# Stochastic model for scale-free networks with cutoffs

Tiago Simas[*]

*Cognitive Science Program, Indiana University, Bloomington, Indiana 47406, USA*

Luis M. Rocha[†]

*School of Informatics and Cognitive Science Program, Indiana University, Bloomington, Indiana 47406, USA*
*and Instituto Gulbenkian de Ciencia, Oeiras, Portugal*

We propose and analyze a stochastic model which explains, analytically, the cutoff behavior of real scale-free networks previously modeled computationally by Amaral *et al.* [Proc. Natl. Acad. Sci. U.S.A. **97**, 11149 (2000)] and others. We present a mathematical model that can explain several existing computational scale-free network generation models. This yields a theoretical basis to understand cutoff behavior in complex networks, previously treated only with simulations using distinct models. Therefore, ours is an integrative approach that unifies the existing literature on cutoff behavior in scale-free networks. Furthermore, our mathematical model allows us to reach conclusions not hitherto possible with computational models: the ability to predict the equilibrium point of active vertices and to relate the growth of networks with the probability of aging. We also discuss how our model introduces a useful way to classify scale free behavior of complex networks.

## I. INTRODUCTION

In the last decade, much work has been done to understand the general mechanisms that influence the growth and dynamics of complex networks. Complex networks have been studied by mathematicians, social scientists, physicists, and others. Perhaps the two most influential contributions are the small world phenomenon proposed by Watts and Strogatz [1] and the preferential attachment mechanism behind scale-free networks proposed by Barabasi and Albert [2]. Because of the pervasiveness of both the small-world phenomenon and scale-free networks in nature and society, there has been extraordinary interest in the study of their structure and dynamics [3–5]. The study of complex networks has been applied to better understand the Internet [3], the World Wide Web [6], protein-protein interaction networks [7], metabolic networks [8,9], and many other natural networks.

We present and analyze a mathematical model which analytically explains the cutoff behavior of real scale free networks observed in the computational model of Amaral *et al.* [10]. We also compare our model with the previous work of Mossa *et al.* [11].

The Amaral *et al.* model is not the only proposed explanation for the cutoffs in a power law degree distribution. Other models such as those in Refs. [11–13], based on finite size effects, have been proposed. However, the Amaral *et al.* model is among the simplest models and therefore amenable to analysis.

The paper is organized as follows. We begin in Sec. II with an overview of the main models of scale-free networks in the literature, including the model of Amaral *et al.* [10] which was used to explain, via simulations, the cutoff behavior of real scale-free network degree distributions. In Sec. III we present the central contribution of this paper which is a stochastic model that analytically explains the cutoff behavior of scale-free networks observed in the simulations of Amaral *et al.* [10]. In Secs. IV and V we discuss our results and their implications for the study of complex networks in general and scale-free behavior, in particular.

## II. SCALE FREE NETWORKS

We conceptualize a network as a graph $G=(V,E)$, where $V$ is a set of vertices (or nodes) $v_i$ and $E$ is a set of edges $e_{i,j}$ which represent a connection between vertices $v_i$ and $v_j$; if the graph is directed $e_{j,i}$ is not necessarily equal to $e_{i,j}$. The degree (or valency) $k_i$ of a vertex $v_i$ is the number of connected vertices (incident edges) to $v_i$. In a directed graph, the indegree $k_i^+$ of a vertex $v_i$ is the number of edges $e_{j,i}$ terminating at $v_i$, and the outdegree $k_i^-$ of a vertex $v_i$ is the number of edges $e_{i,j}$ originating at $v_i$. From this point on, unless otherwise specified, in the case of directed graphs, we will use degree ($k_i$) to mean indegree ($k_i^+$).

It is convenient to characterize large graphs by their degree distribution, which is the distribution of the probability that the degree of a randomly chosen vertex is $k$ [3]. A power law distribution is a distribution that follows the relation

$$P(k) \simeq ak^{-\gamma},$$

where $\gamma$ and $a$ are constants. Newman [14] defines a scale-free network as a graph whose degree distribution follows a power law.

### A. The Barabasi-Albert model

Given an initial connected network (or graph) $G$ with $n_0$ vertices, generally a small network, the Barabasi-Albert model (BAM) [2] is based on the following axioms.

*Axiom 1 (Growth).* A new vertex $v_g$ is added to $G$ at each time step.

---
[*]tdesimas@indiana.edu
[†]rocha@indiana.edu

*Axiom 2 (Preferential attachment)*. An edge $e_{g,i}$ between $v_g$ and $m \leq n_0$ vertices $v_i$ is created at each time step with probability

$$\Pi(e_{g,i}) = \frac{k_i}{\Sigma_j k_j},$$

where $k_i$ is the degree of vertex $v_i$ in the previous time step and $\Sigma_j k_j$ is the total sum of the degree of every vertex in the network in the previous time step. In other words, the preferential attachment axiom biases the generation of new edges towards vertices with higher degrees. With these considerations and the evolution equation

$$\frac{\partial k_i}{\partial t} = m\Pi(e_{g,i}), \tag{1}$$

where, in our case, the constant $m$ is the rate of edges we are introducing each time step, Barabasi and Albert [2] have shown that the model generates a power law distribution, which is independent of time:

$$P(k) \propto k^{-3}.$$

The growth and preferential attachment axioms implement the mechanism known as "the rich gets richer." This mechanism can be generalized in many ways, which are beyond the scope of this paper, for an overview see Ref. [3].

### B. The Amaral *et al.* cutoff model

Amaral *et al.* [10] noticed that in several real networks the power law describing the degree distribution is truncated (or cutoff) for vertices with large degrees. In other words, the number of highly connected vertices is smaller than expected from the preferential attachment model. Several mechanisms can be behind this behavior. In particular, Amaral *et al.* proposed two alternative mechanisms, which interact with the two axioms of the BAM: aging of the vertices and cost of adding edges to the vertices. Both mechanisms produce a power law truncation, i.e., a cutoff in the power law degree distribution. Each alternative mechanism proposed by Amaral *et al.* [10] is defined by an additional axiom to the BAM axioms.

*Axiom 3a (Aging of the vertices)*. At each time step every vertex may become inactive with a constant probability of aging $p$.

An inactive vertex and its edges are still present in the network but it is not allowed to receive more edges. The other mechanism, cost of adding edges to vertices, is similarly implemented by an alternative third axiom.

*Axiom 3b (Cost of adding edges to vertices)*. Each vertex has a limit capacity $k_c$ of edges that it can support. After this threshold a vertex becomes inactive.

Axiom 3b leads to networks whose degree distribution follows the power laws obtained via BAM, except that it observes a spike at $k_c$, followed by an abrupt and unrealistic cutoff. Therefore, the model obtained by axiom 3b is not as realistic and interesting as the one obtained via axiom 3a aging of the vertices, which is the only one we discuss from now on.

Amaral *et al.* [10] have shown with simulations that the BAM with axiom 3a leads to a truncation or a cutoff of the expected power law degree distribution for several probabilities $p$—the behavior observed in many real networks. This truncation is more prominent with higher values of $p$. However, the simulations of Amaral *et al.* do not allow us to determine precisely the ranges of values of $p$ which allow the network to grow. We also, for instance, do not have a precise notion of how the vertices become inactive, or how many vertices are expected to be active in the network at a given time for various values of $p$.

In conclusion, because a mathematical analysis of this model has not been offered, we have not been able to answer such questions. We do know that the mechanism of aging of vertices leads to a behavior similar to real-world networks, and therefore could be the mechanism leading to the questions raised above. Thus, it is important to understand this mechanism in a more analytical manner, which we pursue in this article.

## III. PREFERENTIAL ATTACHMENT WITH VERTEX AGING

### A. Stochastic model

As we discussed in the previous section, the preferential attachment with vertex aging (PAVA) model of Amaral *et al.* [10], is based on three axioms: growth, preferential attachment, and aging of the vertices. In this section we propose a stochastic theoretical model (STM) to study PAVA analytically.

Let us first analyze how the vertices (nodes) become inactive. This is a fundamental piece of the analysis. We start with a core, fully connected network of $x_0$ vertices. Notice that at each time step axiom 3a is equivalent to computing $x(t)$ Bernoulli trials, one for each vertex, where $x(t)$ is the number of vertices at time $t$, and $p$ is the probability that a vertex becomes inactive. The probability of $l$ vertices remaining active after $x(t)$ independent Bernoulli trials is given by the binomial probability distribution

$$P(l,t) = \binom{x(t)}{l}(1-p)^l p^{[x(t)-l]}. \tag{2}$$

Therefore, the dynamics of a network can be expressed by the following stochastic map, where for convenience $x(t) = x_t$ now represents the mean number of vertices at time $t$:

$$x_{t+1} = x_t + \alpha - px_t, \tag{3}$$

where $\alpha$ the number of vertices we introduce at each time step. Because at each time step we perform $x_t$ Bernoulli trials, and introduce $\alpha$ vertices, there are $x_t + \alpha$ vertices in the next time step minus the ones that become inactive; the mean value of which for the binomial distribution is $px_t$ [15]. We can rearrange the terms and write the map in the following way:

$$x_{t+1} = (1-p)x_t + \alpha. \tag{4}$$

The equilibrium points for this map, which refer to the situations when the network retains the same mean number of vertices from iteration to iteration, can be identified by solving the equation

$$(1-p)x_t + \alpha = x_t, \tag{5}$$

which results in the unique equilibrium point $\bar{x}_t = \frac{\alpha}{p}$ that is asymptotically stable when $\|f'(\bar{x})\| < 1$, where the first derivative of the map is given by

$$f(x) = (1-p)x + \alpha \Rightarrow f'(x) = 1 - p.$$

Therefore, the unique equilibrium point $\bar{x}_t = \frac{\alpha}{p}$ is asymptotically stable for $p > 0$. Interestingly, when $p = 0$ our stochastic map yields the pure BAM of Sec. II A. In this case the dynamical system does not have an equilibrium point and it diverges, i.e., the network keeps growing in size. When $p = 1$, the system, of course, does not grow since all vertices become immediately inactive. Finally, when $0 < p < 1$ the system observes the single asymptotically stable equilibrium point $\bar{x}_t = \frac{\alpha}{p}$, which depends on the value of $p$.

The equilibrium behavior of our STM can be better appreciated when we look at the solution of our stochastic map [Eq. (3)] [16]

$$x_t = \begin{cases} x_0 + \alpha t & \text{if } p = 0, \\ \left(x_0 - \frac{\alpha}{p}\right)(1-p)^t + \frac{\alpha}{p} & \text{if } p \neq 0, \end{cases}$$

$$t = 0, 1, 2, \ldots. \tag{6}$$

Because $(1-p) < 1$ for the second condition ($p \neq 0$) we can see that the dynamical system converges to the asymptotically stable equilibrium point $\bar{x} = \frac{\alpha}{p}$. In other words, after a transient the dynamical system converges to a network with a fixed mean number of active vertices and the system remains in that state forever. This transient can be estimated as the time it takes for $(x_0 - \frac{\alpha}{p})(1-p)^t$ to become sufficiently small, which can be better appreciated with a little manipulation of this expression:

$$\left(x_0 - \frac{\alpha}{p}\right)(1-p)^t = Ae^{t\ln(1-p)} = Ae^{-t/t_0}, \tag{7}$$

where

$$A \equiv \left(x_0 - \frac{\alpha}{p}\right),$$

$$t_0 \equiv -\frac{1}{\ln(1-p)}.$$

Now, because the map in our model is stochastic there is variation about the equilibrium point. In our model [Eq. (3)], $x_t$ is a binomial variable and for large enough $t$ we can approximate it by a normal distribution [15] and study its variation. We assume the ergodic hypothesis is true, therefore the statistical mean of $x_t$, which we denote as $\langle x_t \rangle$ is equal to $\bar{x}_t = \frac{\alpha}{p}$, the equilibrium point given by Eq. (6). In our simulation section we validate this assumption. The variations can then be studied by solving the equation for variance

$$\sigma^2 = \langle x_t^2 \rangle - \langle x_t \rangle^2. \tag{8}$$

From Eqs. (3) and (6) and from the ergodic hypothesis we have for the statistical mean

$$\mu = \langle x_{t+1} \rangle = \langle x_t \rangle = \frac{\alpha}{p}. \tag{9}$$

Extending $x_t$ in Eq. (3) to real values to make the approximation to the normal distribution feasible, and substituting this equation and Eq. (9) into Eq. (8),

$$\sigma^2 = \langle [(1-p)x + \alpha]^2 \rangle - \frac{\alpha^2}{p^2} \tag{10}$$

and assuming a normal distribution as discussed above we reach the following expression:

$$\sigma^2 = \frac{1}{\sigma\sqrt{2\pi}} \int_{-\infty}^{\infty} [(1-p)x + \alpha]^2 e^{-(x-\mu)^2/2\sigma^2} dx - \frac{\alpha^2}{p^2} \tag{11}$$

solving Eq. (11) in order of $\sigma$ for $t > t_0$ [Eq. (7)] we obtain

$$\sigma = \sqrt{\frac{1}{1-(1-p)^2}}. \tag{12}$$

For large $t$ Eqs. (9) and (12) define our stochastic variable $x_t$ as a normal variable.

### B. Exponential decay

Let us now take a closer look at axiom 3$a$. According to this axiom a vertex at each time step may become inactive with a constant probability $p$. Each vertex follows a binomial distribution which is equivalent to a random walk process. In this case, it is as if each vertex is trying to give $z$ steps in a maximum of $r$ steps all in the same direction—where $r$ can be interpreted as the maximum iterations $t$. If a vertex changes the direction of its step it becomes inactive. This can be expressed by the following probability function:

$$P(Z = z) = \frac{\binom{r-z}{qr-z}}{\binom{r}{qr}}, \tag{13}$$

where $q = 1 - p$ is the probability that the vertex succeeds and remains active for the next step. This equation can be simplified in the following form:

$$P(Z = z) = \frac{(r-z)!(qr)!(r-qr)!}{(qr-z)!(r-qr)!r!},$$

$$P(Z = z) = \frac{\prod_{k=0}^{z-1}(qr-k)}{\prod_{k=0}^{z-1}(r-k)}. \tag{14}$$

For a given $r = 10\,000$ and $p = 0.1$, we obtain the distribution presented in Fig. 1.

We can see that the respective exponential decay is independent of $r$, by taking the limit when $r \gg 1$:

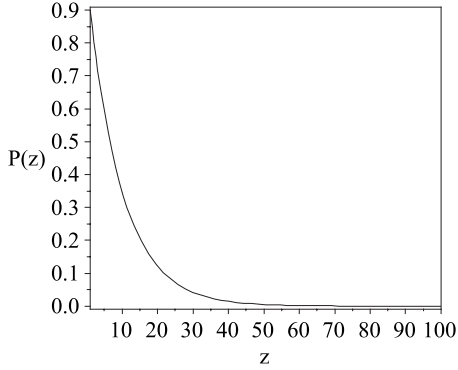

FIG. 1. $P(Z=z)$ distribution.

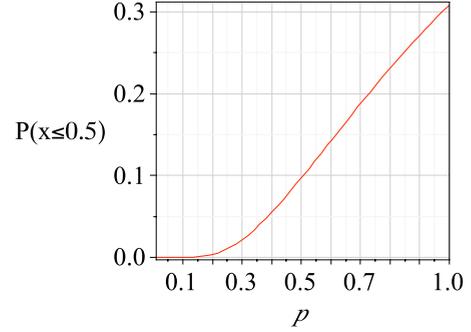

FIG. 2. (Color online) $P(x \leq 0.5)$ versus $p$ (probability of a vertex getting inactive) for $\alpha=1$.

$$P(Z=z) = \frac{\prod_{k=0}^{z-1}(qr-k)}{\prod_{k=0}^{z-1}(r-k)} \underset{r \gg 1}{\cong} \frac{(qr)^{z-1}}{r^{z-1}} = q^{z-1}. \quad (15)$$

In this equation we can see that the binomial experiment of a vertex gets inactive in an exponential decay.

### C. Exponential decay for the degree distribution

It was found in Ref. [17] that for networks where the number of active vertices is a subset of the total of vertices in the network, the power law degree distribution presents a truncation

$$P(K=k) = C(a_m + b_m) k^{-(a_m+1)} e^{-b_m k}, \quad (16)$$

where $m$ is the number of active vertices and $a_m$, $b_m$ are parameters in function of $m$, and $k$ the degree. In our case we are in that situation, where the mean value of vertices active can be represented by $m$. In the next subsection we show how the STM can be validated by simulations of the PAVA model.

### D. Network stop growing estimation

Because the STM is stochastic there is a probability that a network eventually will stop growing. From the results in Secs. III A and III B, we can estimate the probability for which a network will not grow. We have seen that we can approximate $x_t$ as a normal distribution with mean and standard deviation given, respectively, by Eqs. (9) and (12). The next equation estimates the probability of variations (fluctuations) on $x$ such that the network stops growing

$$P(x \leq 0.5) = \Phi\left(\frac{0.5 - \mu}{\sigma}\right), \quad (17)$$

where $\Phi$ is the cumulative probability function of a normal variable with mean $\mu$ and standard deviation $\sigma$—which depends exclusively on the values of $\alpha$ and $p$. We have chosen $x \leq 0.5$ in order to compensate the discrete extension of $x_t$ to real values $x$. Moreover, this probability is the probability that a given network will die in $t_{\text{die}}$ steps. This is exactly the idea behind Eq. (15), but now instead of a given vertex sur-vival probability, we are considering the probability of the network survival given by $w = 1 - P(x \leq 0.5)$, after $t$ time steps and $s$ trials, and for large $s$ we have as before

$$P(T=t) = \frac{\prod_{k=0}^{t-1}(ws-k)}{\prod_{k=0}^{t-1}(s-k)} \underset{r \gg 1}{\cong} \frac{(ws)^{t-1}}{s^{t-1}} = w^{t-1} = Be^{-t/k_c}, \quad (18)$$

with

$$B \equiv \left(\frac{1}{w}\right),$$

$$k_c \equiv -\frac{1}{\ln(w)}.$$

In Fig. 2 we plot the probability for which a network will stop growing after $t$ steps versus the probability of a vertex getting inactive. Most network trials will stop growing after $k_c$. With this we can estimate the number of steps $t_{\text{die}}$ for which the network stops growing. For example, for a probability of inactiveness $p=0.2$ we have $P(x \leq 0.5) = 0.0035$, which means we will have $k_c = 288$. In this case with $p=0.2$ most of the networks will stop growing after $t > 288$.

### E. Simulations

In Table I, we can see the results that we have obtained comparing PAVA simulations with our analytic STM. First a confirmation of the system's stable equilibrium point for each $p$. For the PAVA model we performed eleven simulations for each probability $p$, from $p=0.1$ to $=0.01$, for a network with 10 000 vertices and an $\alpha=1$—eleven simulations is usually considered the minimum experiments. In any case as it can be seen in Table I eleven simulations yielded quite accurate results. The choice for $p$ was based on results obtained in Fig. 2, where we can see for values of $p \leq 0.1$ that the probability of the network not growing is essentially null.

TABLE I. Comparison between PAVA and STM models for $\alpha=1$ and $t=10\,000$.

|  | PAVA[a] | STM |
|---|---|---|
| $p$ | $\bar{x} \pm \sigma$ | $\bar{x} \pm \sigma$ |
| 0.1 | $10.01 \pm 2.18$ | $10.00 \pm 2.29$ |
| 0.08 | $12.49 \pm 2.47$ | $12.50 \pm 2.55$ |
| 0.06 | $16.62 \pm 2.84$ | $16.67 \pm 2.93$ |
| 0.05 | $19.50 \pm 3.10$ | $20.00 \pm 3.20$ |
| 0.03 | $33.56 \pm 4.07$ | $33.33 \pm 4.11$ |
| 0.01 | $99.75 \pm 7.01$ | $100.00 \pm 7.09$ |

[a]Measured after a transient period $t_0$ calculated after 11 simulations.

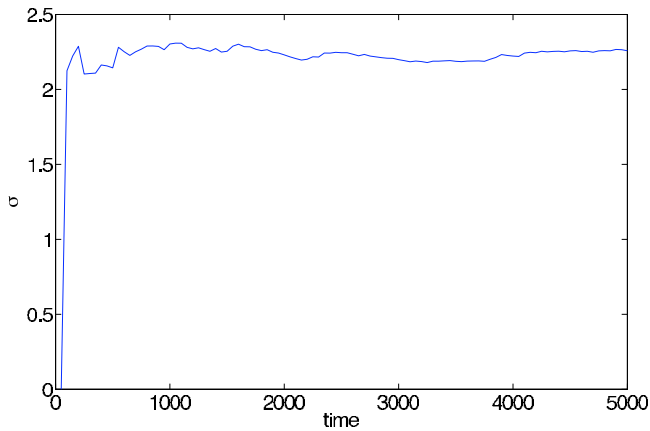

FIG. 3. (Color online) Evolution of the standard deviation of $x_t$ for $p=0.1$ and $\alpha=1$ with time.

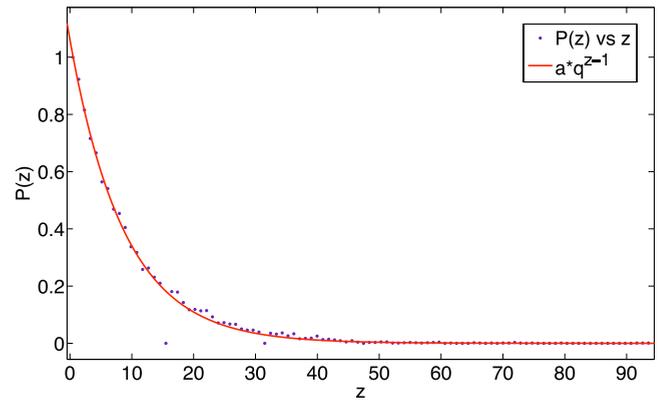

FIG. 4. (Color online) The simulation results for $P(Z=z)$ and curve fitting with an exponential $aq^{z-1}$.

In Table I we see that $\bar{x}$ is extremely well predicted by the STM, since values follow the same tendency as well as the standard deviation obtained by the PAVA simulations. In Fig. 3 we can see that the standard deviation stabilizes after a transient period of time as was predicted by Eq. (7).

In Table II we compare the number of iterations for which most of the networks stop growing for both PAVA simulations and the STM. The choice for $p$ was based on results obtained in Fig. 2. The STM values were estimated by using Eq. (18). To estimate the cutoff point of the exponential decay in PAVA we have fit with a 95% confidence an exponential function $ae^{-bt}$, to the experimental data. The estimated parameter $b$, allow us to compute $t_c = \frac{1}{b}$, which is the exponential cutoff. The comparison between PAVA and STM is made between $t_c$ and $k_c$. There are some fluctuations but both follow the same tendency, which shows that eventually the network will stop growing after $t_c$ or $k_c$ iterations.

From the results summarized in Tables I and II it can be concluded that the number of active vertices observed by PAVA simulations follows the process described by the dynamic map inherent in the STM; in other words the aging of vertices process is a binomial random process.

TABLE II. All vertex getting inactive after $t$ iterations according to the probability $p$ with $\alpha=1$. $t_c$ is the time step for which the PAVA network stops growing, $k_c = -\frac{1}{\ln[1-P(x \leq 0.5)]}$ is the cutoff point, and $P(x \leq 0.5)$ the theoretical probability for which a network will stop growing.

|   | PAVA | STM |
|---|------|-----|
| $p$ | $t_c$ | $k_c$ |
| 0.2 | 374 | 288 |
| 0.3 | 33 | 46 |
| 0.4 | 11 | 18 |

The PAVA and STM probability distribution $P(Z=z)$ for the number of steps a vertex succeeds without getting inactive is shown in Fig. 4 for $r=10\,000$ vertices and $p=0.1$. In these figures we see the experimental probability distribution (dots) and the STM distribution fits Eq. (15) well, $q^{z-1}$, where $q=1-p=0.9$. The results of the regression are an estimated $q=0.89$ for 95% of confidence with sum squared error, SSE=0.04, and $R^2=0.99$, and root mean square errors, RMSE=0.02.

In Fig. 5 we see that the degree distribution cutoff of the PAVA simulations does not change with the size as already observed by Amaral et al. [10]. Similar results are observed for other $p$. The power law exponent $\gamma$ does not change significantly; in the case of Fig. 5, the $\gamma$ values are around 2.6.

In Fig. 6, also as expected, the degree distribution cutoff point of the PAVA simulations decreases inversely with the probability of inactiveness $p$ as already observed by Amaral et al. [10]. Also, the the power law exponent $\gamma$ does not change significantly with $p$.

These results show that (a) the network after a transient period of time $t_0$ reaches an equilibrium number of active vertices, (b) the network will eventually stop growing after $t$ iterations according to the probability of inactiveness. According to all these results we can conclude that our STM is a good analytical model of the generation processes of PAVA, namely, the aging of vertices process.

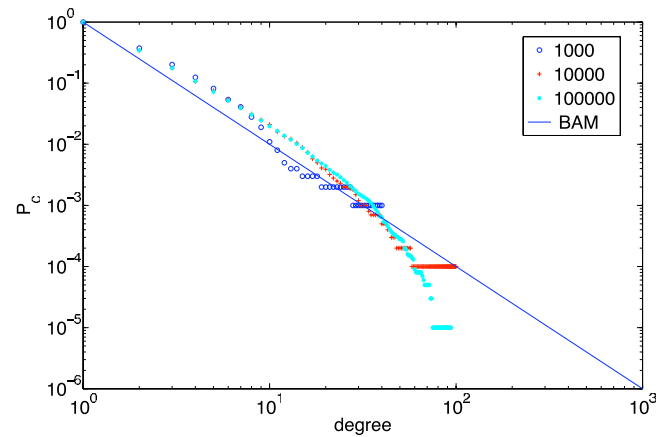

FIG. 5. (Color online) Cumulative degree distribution for network sizes 1000, 10 000, and 100 000, with $\alpha=1$ and $p=0.05$. Also in solid we plot the BAM.

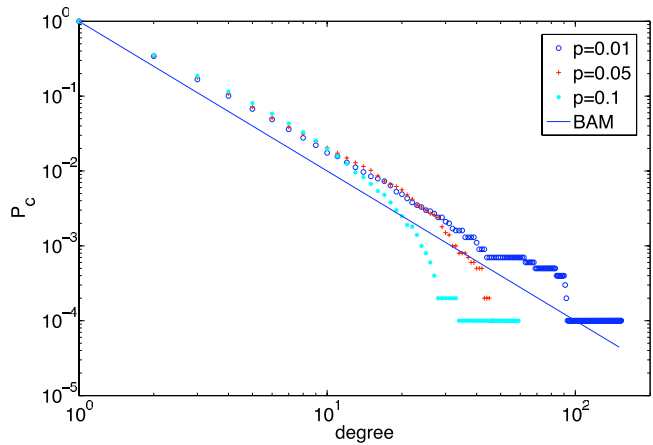

FIG. 6. (Color online) Cumulative degree distribution for probability of inactiveness $p=0.1$, $p=0.05$, and $p=0.01$, with $\alpha=1$ and size of the network=10 000. Also in solid we plot the BAM.

## IV. DISCUSSION

We have seen in previous sections that the PAVA model follows a binomial distribution and can be described by a discrete dynamical system. This dynamical system has a stable equilibrium point and two extreme behaviors, the pure random walk, when there is only one active vertex, and at the other extreme the Barabasi-Albert behavior, when we have all vertices active. Moreover, in addition to the simulations results made by Amaral *et al.* [10] (PAVA) our STM was also able to (a) predict the equilibrium point of active vertices and (b) relate the growth of the network (size) with the probability of aging.

The PAVA simulations are an extension of the BAM and is useful for the study of several real complex networks, such as the Actor network and Scientific Citations networks [18]. These networks are characterized by vertex dying over time. In the case of the actors network, an actor during a period of time plays in several movies with other actors and then they become inactive (die or retired). However, the influence of inactive actors still participates on the statistics of the network without removing the actor from the network. The same kind of reasoning can be applied to the network of citations, where a given paper becomes inactive (obsolete) after a given period of time but remains on the network.

Another way of interpreting the PAVA simulations of Amaral *et al.* [10] and observe the relation between this generation process with others in the literature, is if we look at the network from the perspective of a new vertex. In PAVA a vertex gets inactive with a probability $p$ as a result of a binomial process. Probabilistically equivalent to this is if a new vertex, at each time step, just sees a limited number of random vertices, the active vertices, which are themselves limited by the equilibrium point reached by the network after a transient period of time. This can be seen as an information filtering done by each new vertex regarding the whole network. It is not possible for a new vertex to have a full knowledge of the entire network, it just has a partial knowledge. Therefore we can say the stochastic map defined in Eq. (6), represents the knowledge that a new vertex has about the entire network.

If $p=0$ a new vertex has full knowledge of the network and in this case we are in one extreme, the purely Barabasi-Albert model. In the other extreme if we have a certain maximum $p_{max}<1$ where the equilibrium point is $\bar{x}=1$, only one vertex is active. In this case the new vertex does not have any knowledge at all; it just connects to the other vertices in a random way. The intermediate case happens when $p<p_{max}$ and $\bar{x}>1$. In this case each new vertex has partial knowledge of the network. Therefore, some scale-free networks range between two situations; absolute knowledge of the network, the Barabasi-Albert model [3], and complete ignorance of the network, pure random process. The PAVA simulations of Amaral *et al.* [10] and obviously its STM formalization, seems to be a simpler model that could explain, as a first approximation, the general dynamic mechanisms behind scale-free networks between these two extremes. The parameter $p$ measures how each vertex has complete knowledge of the network or complete ignorance.

## V. CONCLUSIONS

In this work we have presented a stochastic theoretical model as a mathematical explanation of the PAVA model of Amaral *et al.* [10]. We believe this work can provide a simple explanation for the dynamics of some scale-free networks and through this knowledge, obtain a better understanding of how these scale-free networks can emerge. As we described in the Introduction, the field of complex networks is an interdisciplinary field. Therefore a better understanding of the mechanisms behind complex networks can improve the understanding behind certain problems in areas such as the Internet, World Wide Web, Neural Networks, Chemical Networks, Social Networks, and so on.

## ACKNOWLEDGMENTS

This work was partially supported by Portuguese Foundation of Science and Technology Grant No. SFRH/BD/6265/2001. We would also like to thank for various discussions Alessandro Vespignani, Marc Berthelemy, Artemy Kolchinsky, Alessandro Flammini, and Rita Ribeiro.